\documentstyle[12pt]{article}

\begin{document}
\begin{titlepage}
\title{ NEUTRINO EMISSIVITY OF DEGENERATE \\
 QUARK MATTER}
\author{ Sanjay K. Ghosh, S. C. Phatak and Pradip K. Sahu\\
Institute of Physics, Bhubaneswar-751005, India}
\maketitle
\thispagestyle{empty}
\begin{abstract}
An exact numerical calculation of neutrino emissivity of two and
three flavour quark matter have been carried out. We find that the
neutrino emissivity obtained from the Iwamoto formula is in qualitative
agreement with our calculation for two flavour quark matter. For
three flavour case, on the other hand we find that the Iwamoto
formula overestimates the numerical values by $\sim$2 orders of
magnitude or more for d decay and agrees with the s decay results
within a factor of 3-4. The dependence of the emissivity on
temperature, strong coupling constant and baryon density is also
quite different from the Iwamoto formula.

\end{abstract}
\smallskip
{\it{Subject headings}}: Elementary particles - neutrinos - dense stars :
 neutron
\end{titlepage}
\eject

\newpage
\hspace {0.5in} It has been conjectured that dense stars may
consist of quark matter or quark matter core with neutron matter
outside  ( Bahcall $\&$ Wolf 1965, Burrow 1980, Freedman $\&$ McLerran 1978,
Maxwell et al. 1977, Witten 1984 ). Although  theoretical
understanding of the properties of quark matter is not yet available,
various quark models have been used to calculate  the
equation of state of the quark matter and determine the properties of
quark stars  ( Farhi $\&$ Jaffe 1984,
Ghosh $\&$ Sahu 1993, Goyal $\&$ Anand 1990, Haensel et al. 1986
). Unfortunately, it is found that the properties of quark
stars,
such as surface gravitational redshift z,moment of inertia I,
maximum mass M, radius R and pulsar periods P, are not
significantly different compared to those of neutron stars.
Therefore it is difficult to distinguish one from the other
observationally.

    On the other hand Iwamoto ( Iwamoto 1980, 1982 )  has proposed that
neutrino
emissivity  ( $\epsilon$ )  could play a significant role in
distinguishing between quark and neutron stars because it
differs by orders of magnitude for the two. Particularly
$\epsilon$ for quark stars is larger by 6-7  orders of
magnitude than neutron stars which could lead to faster cooling
rate for quark stars, thus reducing their surface temperature. There
are, however, a number of other
mechanisms ( pion condensation ( Brown 1977 ) and modified URCA
processes ( Friman $\&$ Maxwell 1979  ) )
proposed to increase $\epsilon$ for neutron matter.

    Iwamoto ( 1980 )  has derived the formula for $\epsilon$
using apparently reasonable approximations and this formula has
been widely used ( Alcock et al. 1986, Datta et al. 1988, Duncan et
al. 1983, Goyal $\&$ Anand 1990 ) to calculate $\epsilon$ for
two and three flavour quark matter. According to his formula
$\epsilon$ is proportional to baryon density  ( $n_B$ ) , strong coupling
constant  ( $\alpha_c$ )  and sixth power of temperature (T) for d
quark decay. For s quark decay T dependence of $\epsilon$ is
same as that for d decay. Furthermore
his results show that electron and quark masses have negligible
contribution to $\epsilon$ and s quark decay  ( in
case of three flavour quark matter ) plays a rather insignificant
role.

    In the present letter we want to report an exact numerical
calculation of $\epsilon$ and a comparison of our results
with the Iwamoto formula. Our results show that the Iwamoto
formula is in qualitative agreement with our calculation for two
flavour quark matter, although it yields $\epsilon$ which is 3-4
times larger. However we find that the Iwamoto formula overestimates
$\epsilon$ by about 2 orders of magnitude or more
for three flavour quark matter and hence it should not be used
for three flavour quark matter. In the following, we first discuss
the equilibrium composition of matter then we
sketch the  calculation of $\epsilon$ and present the results.

    The simplest possible processes  for neutrino emission in two
flavour degenerate quark matter are direct $\beta$ decay
reactions
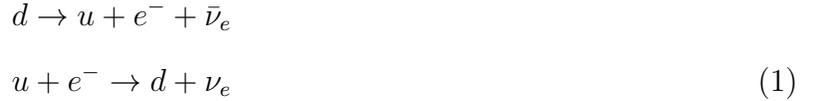
\begin{eqnarray}
d \rightarrow u + e^{-} +\bar \nu_e \nonumber \\
u+ e^- \rightarrow d+ \nu_e
\end{eqnarray}
The chemical equilibrium and charge neutrality implies
\begin{equation}
\mu_d = \mu_u + \mu_e   (\mu_{\nu_e}=\mu_{\bar\nu_e}=0)
\end{equation}
and
\begin{equation}
 2 n_u - n_d -3 n_e = 0
\end{equation}
The baryon density is defined as $n_B = ( n_u + n_d )/3 $
where $n_i = g.p_{{F}_{i}}^{3}/(6 \pi^2)$ is the particle number
density. Degeneracy factor g is 6 for quarks and 2 for electron.

For three flavour degenerate quark matter the neutrino
emission occurs via d as well as s decay. The s decay
reactions are
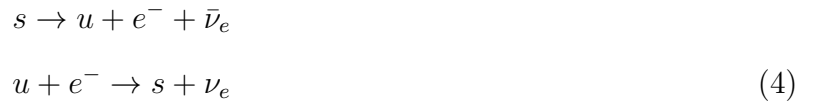
\begin{eqnarray}
s \rightarrow u + e^- + \bar\nu_e \nonumber \\
u + e^- \rightarrow s + \nu_e
\end{eqnarray}
 From equations (1), (2) and (4), the chemical equilibrium and charge
neutrality gives
\begin{eqnarray}
\mu_s = \mu_u + \mu_e (\mu_{\nu_e}=\mu_{\bar\nu_e}=0) \nonumber \\
\mu_d = \mu_s
\end{eqnarray}
and
\begin{equation}
 2n_u - n_d -n_s -3n_e = 0
\end{equation}

Baryon density is defined as $n_B=(n_u+n_d+n_s)/3$.
Thus, given the baryon density, quark and electron number
densities and chemical potentials are calculated using the
constraints above. These are then used to evaluate neutrino
emissivity.

      Neutrino emissivity $\epsilon_{d(s)}$ for d (s) decay is given
by ( Iwamoto 1980,1982 )
\begin{eqnarray}
\epsilon_{d(s)} &=&  A_{d(s)}\int d^3 p_{d(s)} d^3 p_u d^3 p_e d^3
p_{\nu} {( p_{d(s)}
. p_{\nu} ) ( p_u . p_e ) \over { E_u E_{d(s)} E_e }}
\nonumber\\  &  &
\times \delta^{4}(p_{d(s)}-p_u-p_e-p_{\nu}) n(\vec p_{d(s)}) [1-n(\vec
p_u)][1-n(\vec
p_e)]
\end{eqnarray}

where $p_{d(s)}$, $p_u$, $p_e$, $p_\nu$ are the four momenta of d(or
s), u, e, $\nu$; $n(\vec p_i)$ (i= u,d (s), e) is the Fermi
distribution given by
\begin{equation}
n(\vec p_i) = {1\over{ e^{\beta (E_i - \mu_i)} +1 }}
\end{equation}
and
\begin{eqnarray}
A_d = {24 G^2 cos^2\theta_c\over{(2 \pi )^8}}\\
A_s = {24 G^2 sin^2\theta_c\over{(2 \pi )^8}}
\end{eqnarray}
where G is Weak coupling constant and $\theta_c$ is Cabibbo
angle, $E_i$ is the energy and $p_i$ is the momentum.
In terms of Fermi momentum $p_F$ the quark chemical potential
$\mu$  ( upto lowest order in $\alpha_c$ )  is  (  Baym $\&$ Chin 1976  )
\begin{equation}
\mu = [ {\eta \over x} + {8\alpha_c \over { 3 \pi}} ( 1- {3\over
{x \eta}} ln(x+\eta))]p_{F}
\end{equation}
where $x\equiv p_{F}/m$ and $\eta \equiv \sqrt(1+x^2)$,
$m$ being the quark mass. For massless quarks eq(11) reduces to
\begin{equation}
\mu=(1+ {{8 \alpha_c}\over{3 \pi}}) p_{F}
\end{equation}
Since in our calculation we have taken u, d quark masses to be zero
and s quark mass to be nonzero ( 150 - 200 MeV),
the energy momentum relation of quarks is approximated by
eq(11) for s quarks and eq(12) for u and d quarks, that is, $\mu$
is replaced by $E$ and $p_F$ is replaced by p in above eq(11-12).
This is reasonable as only the energies and momenta close to Fermi
surface contribute to the matrix element.
 For massless electrons $E_e = p_e$ and the
energy delta function is used to perform $p_e$ integral. This gives
\begin{equation}
p_e= {{(E_{d(s)}-E_u)^2- (p_{d(s)}-p_u)^2- 2 p_{d(s)} p_u
(1-cos\theta_{u})}\over{ (2 p_u cos\theta_{ue}+ 2 p_{d(s)}
cos\theta_e+ 2 E_{d(s)}- 2 E_u)}}
\end{equation}
where $\theta_{u}$, $\theta_{e}$ and $\theta_{ue}$ are the
angles between d and u, d and e and u and e respectively.
The momentum delta function in  eq(7) is used to integrate over the
neutrino momentum $p_\nu$. Of the rest of the integrals $d\Omega_d
d\phi_u$ integral gives $8\pi^2$ and we get

\begin{eqnarray}
\epsilon_{d(s)}&=& 8\pi^2 A_{d(s)} \int p_{d(s)}^2 dp_{d(s)} p_{u}^2 dp_u
p_{e}^{2} dcos\theta_u dcos\theta_e d\phi_e \nonumber \\
& &\times{( p_{d(s)}. p_{\nu} ) ( p_u . p_e ) \over { E_u E_{d(s)} E_e
}} n(\vec p_{d(s)}) [1-n(\vec p_u)][1-n(\vec p_e)]
\end{eqnarray}

In the above expression (eq(14)) limits for momentum integrals are from 0 to
$\infty$, for $cos\theta_u$, $cos\theta_e$ from -1 to +1 and for
$\phi_e$ from 0 to $2\pi$ . However we find that the integrand is a
sharply peaked function about $p_u\sim p_{f}(u)$, $p_d\sim
p_{f}(d)$ (or $p_{s}\sim p_{f}(s)$) , $\phi_e\sim \pi$ and certain
values of $cos\theta_e$ and
$cos\theta_u (\leq 1)$. Thus one must ensure that there are
sufficient number of integration points in this region. This we have
done. We have also ensured the convergence of the integral with
respect to the number of integration points and step length for
integration.

Two and three flavour $\epsilon$,  for different T, $\alpha_c$
and $n_B$  are given in Tables 1 and 2 (a, b) respectively.
Corresponding Fermi momenta  are given in Tables 3 and 4.
For two flavour quark matter $\epsilon_d$ varies as $T^{5.3-5.9}$,
which is slightly lower than the Iwamoto results . In fact the
T variation changes with temperature as well as $\alpha_c$ and
$n_B$. We also find that dependence of $\epsilon$ on $\alpha_c$
and $n_B$ are approximately $\epsilon \propto n_B^{1.03}$ and
$\epsilon \propto \alpha_c^{0.9}$ respectively. These are close
to values obtained  by Iwamoto and others  (  Iwamoto 1982, Alcock et
al. 1986, Datta et al. 1988, Duncan et al. 1983  ) . We find that with
increase in baryon density, electron fraction in two flavour
matter increases, which seems to play a crucial role resulting
in the increase in $\epsilon$. The calculated $\epsilon$ from
the Iwamoto formula has also been displayed in Table 1. A
comparison shows that Iwamoto results are larger by factor of 3
or more with the difference increasing with T and $n_B$. Thus we
find that Iwamoto formula provides a reasonable approximation to
an exact calculation if the difference of a factor of 3-4 is acceptable.

The neutrino emissivity and its
dependence on T, $\alpha_c$ and $n_B$ are entirely different for three
flavour quark matter. For d decay
$\epsilon_d$ varies as $T^{3.6-4.7}$ for $m_s=150 MeV$
and $T^{3.95}$ for $m_s= 200 MeV$. The $\alpha_c$ and $n_B$
dependences also vary in a wide range; for T=0.1 MeV $\epsilon_d
\propto {\alpha_c^{-4}}$ and $\epsilon_d \propto
{n_B^{-4}}$, where as for T= 0.6 MeV $\epsilon_d \propto
{\alpha_c^{-1/2}}$ and $\epsilon_d \propto
{n_B^{-2}}$, $m_s$ being 150 MeV. Unlike the two
flavour case, here electron density decreases with increasing
baryon density. Corresponding $\epsilon_d$ also decreases. The
$\epsilon_d$ obtained using the Iwamoto formula comes out to be much
larger than our values as shown in Table 2a and 2b.
For s decay $\epsilon$ has been found to be much
different when compared with d decay. We find $\epsilon_s \propto
T^{5.00-5.98}$ for $m_s=150 MeV$ and $T^{5.8-6.5}$ for $m_s=200
MeV$. Further increase in $m_s$ results in increase in power of
T beyond 6.
The $\alpha_c$ dependence of $\epsilon_s$ is similar to that of
$\epsilon_d$. One particular change to be noticed is that for
$\alpha_c=0.05$, $\epsilon_s$ decreases with density, whereas for
$\alpha_c=0.1$ it increases with density. This difference in s
decay compared to d decay implies the significant role played
by $m_s$.

	Here we would like to note that the cases where Iwamoto formula
agrees reasonably well with our result, the difference between
$p_{f}(u)+ p_{f}(e)$ and $p_{f}(d)$ ( or $p_{f}(s)$) is much larger
than the temperature. On the other hand, when this difference is
smaller or comparable with the temperature, the Iwamoto formula
overestimates the exact result by orders of magnitude. This behaviour
is also seen when the temperature is larger.

	In order to investigate this further, we have attempted to
evaluate the integral approximately. For this we note that the
integrand peaks at certain values of the integral variables
( $p_{d(s)}(m)$,$p_u(m)$, $cos\theta_u(m)$,$cos\theta_e(m)$ and
$\phi_e(m)$). We find that a reasonable Gaussian approximation for
the integration is

\begin{eqnarray}
\epsilon_{d(s)}&=& C. 8\pi^2 \int dp_{d(s)} dp_u
dcos\theta_u dcos\theta_e d\phi_e
e^{-\alpha_1 ^{2}(p_{d(s)}(m) -p_d)^2}
e^{-\alpha_2 ^{2}(p_{u}(m)
-p_u)^2}\nonumber\\& & \times e^{-\alpha_3 ^{2}(cos\theta_u(m)
-cos\theta_u)^2}e^{-\alpha_4 ^{2}(cos\theta_e(m)
-cos\theta_e)^2}
e^{-\alpha_5 ^{2} (\phi_e(m)-\phi_e)^2}
\end{eqnarray}

The above integration eq(15) can be performed analytically if the
integration limits are from $-\infty$ to $+\infty$ and the result is

\begin{equation}
\epsilon_{d(s)}= C .8\pi^2
\frac{\pi^{5/2}}{\alpha_{1}\alpha_{2}
\alpha_{3}\alpha_{4}\alpha_{5}}
\end{equation}

We find that for both two and three flavour quark matter, the
emissivity from above approximation (eq(16)) are within factor
of 2 of our exact numerical results.
Further we find that for two flavour case $C \propto T $ and
$\alpha_{i}\propto {1\over T}$, so that $\epsilon\propto T^6$
(approximately). For three flavour d decay, C is not linear in T ($\sim
T^{1/2}$) and the dependence of $\alpha_{i}$'s on T is also different.
 Thus, in this case the
temperature dependence obtained after numerical integration differs
significantly from Iwamoto results.

It was first pointed out by Duncan et al. ( 1983 )
that treating electron fraction as a constant  ( $Y_e=0.01$ by
Iwamoto )  is not correct and a detailed treatment of the
equilibrium composition of quark matter is needed. For a fixed
density they solve the chemical equilibrium  and number density
conditions to get the $p_{F}(i)$ and $\mu(i)$. We indeed follow
the same method. Thus the neutrino emissivity denoted by
$\epsilon_{dI(sI)}$ in Table 2 ( a, b ) are essentially the three
flavour $\epsilon$ calculated by  Duncan et al. (  1983 ).

Our calculation shows that for 2-flavour quark matter, $\epsilon$
calculated from the Iwamoto formula agrees qualitatively with the exact
calculation and therefore earlier conclusions ( Alcock et
al. 1986, Datta et al. 1988, Duncan et al. 1983, Iwamoto 1982
 ) regarding cooling rates of 2-flavour quark matter are still valid.
However for 3-flavour quark matter $\epsilon$ is smaller than the
results obtained using the Iwamoto formula by 2-3 orders of magnitude
or more for d decay and factor of 3 or more for s decay. As a
result, $\epsilon$ of three flavour quark matter is
atleast 2 orders of magnitude smaller than that predicted by
the Iwamoto formula. Therefore, the cooling rates of three flavour
quark matter are considerably smaller than the rates estimated
from the Iwamoto formula.

Our calculation shows that $\epsilon$ of two flavour
quark stars is about 3 or more  orders of magnitude larger than that of
three flavour quark stars. Thus, a two flavour quark star will cool
more rapidly than a three flavour quark star. On the other hand the
dynamics of the quark matter suggests that the three flavour quark matter
is most stable ( Witten 1984 )  and a two flavour quark matter
will decay into three flavour quark matter by weak interactions.
We are not aware of calculations where this decay rate has been
calculated. But, as the two to three flavour conversion rate as
well as neutrino cooling involve weak interactions, the two
rates would be comparable.
Furthermore, there are arguments ( Alcock et al. 1986 ) to suggest
that quark stars, when formed, may consist of three flavour quark
matter. Therefore neutrino cooling rates of quark stars may be much
smaller than those suggested from earlier calculations. In this
context, we feel that the knowledge of two to three flavour decay
rate will be useful.

\newpage
\hspace{0.35in}{\bf { REFERENCES}}\hfil\break
\vskip .2in
\noindent  Alcock, C., Farhi, E., $\&$ Olinto, A. 1986, ApJ, 310, 261

\noindent  Bahcall, J. N., $\&$ Wolf, R. A. 1965, Phys. Rev. B,
140,1452

\noindent  Baym, G. $\&$ Chin, S. A. 1976, Nucl. Phys. A, 262, 527

\noindent  Brown, G. E. 1977, Comm. Astrophys. Space. Phys., 7, 67

\noindent  Burrow, A. 1980, Phys. Rev. Lett., 44, 1640

\noindent  Datta, B., Raha, S., $\&$ Sinha, B. 1988, Mod. Phys. Lett.
A, 3, 1385

\noindent  Duncan, R. C., Shapiro, S. L., $\&$ Wasserman, I. 1983,
ApJ, 267, 358

\noindent  Farhi, E., $\&$ Jaffe, R. L. 1984, Phys. Rev. D, 30, 2379

\noindent  Freedman, B., $\&$ McLerran, L. 1978, Phys. Rev. D, 17, 1109

\noindent  Friman, B. L., $\&$ Maxwell, O. V. 1979, ApJ, 232, 541

\noindent  Ghosh, S. K., $\&$ Sahu, P. K. 1993, Int. J. Mod.
Phys. E (in press)

\noindent  Goyal, A., $\&$ Anand, J. D. 1990, Phys. Rev. D, 42, 999

\noindent  Haensel, P., Zdunik, J. L., $\&$ Schaeffer, R. 1986,
Astron. Astrophys., 160, 121

\noindent  Iwamoto, N. 1980, Phys. Lett. Rev., 44, 1196

\noindent  {----------------}   1982, Ann. of Phys., 141, 1

\noindent  Maxwell, O., Brown, G. E., Campbell, D. K., Dashen,\hfil\break
R. F.,$\&$ Manassah, J. T. 1977, ApJ, 216, 77

\noindent  Tsuruta, S. 1979, Phys. Rep., 56, 237

\noindent  Witten, E. 1984, Phys. Rev. D, 30, 272

\newpage
\noindent Table 1.  Neutrino emissivity for two flavour quark matter. Here
$\epsilon_{dI}$ is the emissivity calculated using Iwamoto formula.
\vspace {0.2in}
\begin{center}
\begin{tabular}{|c|c|c|c||c|c|}
\hline
\multicolumn{1}{|c|}{T} &
\multicolumn{1}{|c|}{$n_B$} &
\multicolumn{2}{|c||}{$\alpha_c = 0.1$} &
\multicolumn{2}{|c|}{$\alpha_c = 0.05$} \\
\cline{3-6}
\multicolumn{1}{|c|}{(MeV)} &
\multicolumn{1}{|c|}{$(fm^{-3})$} &
\multicolumn{1}{|c|}{$\epsilon_d$} &
\multicolumn{1}{|c||}{$\epsilon_{dI}$} &
\multicolumn{1}{|c|}{$\epsilon_d$} &
\multicolumn{1}{|c|}{$\epsilon_{dI}$} \\
\multicolumn{1}{|c|}{} &
\multicolumn{1}{|c|}{} &
\multicolumn{1}{|c|}{$(erg/cm^{3}/s)$} &
\multicolumn{1}{|c||}{$(erg/cm^{3}/s)$} &
\multicolumn{1}{|c|}{$(erg/cm^{3}/s)$} &
\multicolumn{1}{|c|}{$(erg/cm^{3}/s)$} \\
\hline
   &0.6&9.93$\times 10^{+25}$&3.01$\times 10^{+26}$&5.21$\times
10^{+25}$&1.45$\times 10^{+26}$ \\
0.1&1.0&1.68$\times 10^{+26}$&5.01$\times
10^{+26}$&8.86$\times10^{+25}$&2.41$\times 10^{+26}$\\
   &1.4&2.35$\times 10^{+26}$&7.02$\times 10^{+26}$&1.25$\times
10^{+26}$&3.38$\times 10^{+26}$\\
\hline
   &0.6&3.35$\times 10^{+29}$&1.23$\times 10^{+30}$&1.15$\times
10^{+29}$&5.93$\times 10^{+29}$ \\
0.4&1.0&5.84$\times 10^{+29}$&2.05$\times 10^{+30}$&2.28$\times
10^{+29}$&9.88$\times 10^{+29}$\\
   &1.4&8.34$\times 10^{+29}$&2.87$\times 10^{+30}$&3.50$\times
10^{+29}$&1.38$\times 10^{+30}$\\
\hline
   &0.6&3.12$\times 10^{+30}$&1.40$\times 10^{+31}$&7.50$\times
10^{+29}$&6.75$\times 10^{+30}$ \\
0.6&1.0&5.77$\times 10^{+30}$&2.34$\times 10^{+31}$&1.62$\times
10^{+30}$&1.12$\times 10^{+31}$\\
   &1.4&8.50$\times 10^{+30}$&3.27$\times 10^{+31}$&2.65$\times
10^{+30}$&1.57$\times 10^{+31}$\\
\hline
\end{tabular}
\end{center}
\newpage
\noindent Table 2a.  Neutrino emissivity for three flavour
matter. Here $\epsilon_{dI}$ and $\epsilon_{sI}$ are the
emissivity calculated using Iwamoto formula.
\vspace {0.1in}
\begin{center}
\begin{tabular}{|c|c|c|c|c|c|c|c|}
\hline
\multicolumn{1}{|c|}{$\alpha_c$} &
\multicolumn{1}{|c|}{$m_s$} &
\multicolumn{1}{|c|}{T} &
\multicolumn{1}{|c|}{$n_B$} &
\multicolumn{1}{|c|}{$\epsilon_d$} &
\multicolumn{1}{|c|}{$\epsilon_{dI}$} &
\multicolumn{1}{|c|}{$\epsilon_s$} &
\multicolumn{1}{|c|}{$\epsilon_{sI}$} \\
\multicolumn{1}{|c|}{} &
\multicolumn{1}{|c|}{(MeV)} &
\multicolumn{1}{|c|}{(MeV)} &
\multicolumn{1}{|c|}{$(fm^{-3})$} &
\multicolumn{1}{|c|}{$(erg/cm^{3}/s)$} &
\multicolumn{1}{|c|}{$(erg/cm^{3}/s)$} &
\multicolumn{1}{|c|}{$(erg/cm^{3}/s)$} &
\multicolumn{1}{|c|}{$(erg/cm^{3}/s)$} \\
\hline
   &     &   &0.6&1.31$\times 10^{+23}$&8.08$\times 10^{+24}$&7.47$\times
10^{+23}$&1.73$\times 10^{+24}$\\
   &     &0.1&1.0&2.11$\times 10^{+22}$&5.73$\times 10^{+24}$&4.81$\times
10^{+23}$&1.23$\times 10^{+24}$\\
   &     &   &1.4&3.46$\times 10^{+21}$&3.56$\times 10^{+24}$&2.34$\times
10^{+23}$&7.60$\times 10^{+23}$\\
\cline{3-8}
   &     &   &0.6&3.19$\times 10^{+25}$&2.31$\times 10^{+28}$&2.23$\times
10^{+27}$&7.10$\times 10^{+27}$\\
   &150.0&0.4&1.0&9.73$\times 10^{+24}$&2.35$\times 10^{+28}$&8.15$\times
10^{+26}$&5.03$\times 10^{+27}$\\
   &     &   &1.4&4.21$\times 10^{+24}$&1.46$\times 10^{+28}$&1.62$\times
10^{+26}$&3.11$\times 10^{+27}$\\
\cline{3-8}
   &     &   &0.6&2.07$\times 10^{+26}$&3.77$\times 10^{+29}$&1.84$\times
10^{+28}$&8.09$\times 10^{+28}$\\
   &     &0.6&1.0&8.48$\times 10^{+25}$&2.67$\times 10^{+29}$&5.02$\times
10^{+27}$&5.74$\times 10^{+28}$\\
0.1&     &   &1.4&4.75$\times 10^{+25}$&1.66$\times 10^{+29}$&8.74$\times
10^{+26}$&3.55$\times 10^{+28}$\\
\cline{2-8}
   &     &   &0.6&2.87$\times 10^{+24}$&2.39$\times 10^{+25}$&2.39$\times
10^{+24}$&5.13$\times 10^{+24}$\\
   &     &0.1&1.0&1.31$\times 10^{+24}$&2.17$\times 10^{+25}$&2.18$\times
10^{+24}$&4.66$\times 10^{+24}$\\
   &     &   &1.4&6.02$\times 10^{+23}$&1.93$\times 10^{+25}$&1.86$\times
10^{+24}$&4.14$\times 10^{+24}$\\
\cline{3-8}
   &     &   &0.6&7.98$\times 10^{+26}$&9.79$\times 10^{+28}$&8.42$\times
10^{+27}$&2.10$\times 10^{+28}$\\
   &200.0&0.4&1.0&2.88$\times 10^{+26}$&8.90$\times 10^{+28}$&7.41$\times
10^{+27}$&1.91$\times 10^{+28}$\\
   &     &   &1.4&1.30$\times 10^{+26}$&7.90$\times 10^{+28}$&6.09$\times
10^{+27}$&1.69$\times 10^{+28}$\\
\cline{3-8}
   &     &   &0.6&3.77$\times 10^{+27}$&1.12$\times 10^{+30}$&9.30$\times
10^{+28}$&2.39$\times 10^{+29}$\\
   &     &0.6&1.0&1.47$\times 10^{+27}$&1.01$\times 10^{+30}$&7.71$\times
10^{+28}$&2.17$\times 10^{+29}$\\
   &     &   &1.4&7.44$\times 10^{+26}$&9.00$\times 10^{+29}$&5.70$\times
10^{+28}$&1.93$\times 10^{+29}$\\
\hline
\end{tabular}
\end{center}
\newpage
\noindent Table 2b. Neutrino emissivity for three flavour
matter. Here $\epsilon_{dI}$ and $\epsilon_{sI}$ are the
emissivity calculated using Iwamoto formula.
\vspace {0.1in}
\begin{center}
\begin{tabular}{|c|c|c|c|c|c|c|c|}
\hline
\multicolumn{1}{|c|}{$\alpha_c$} &
\multicolumn{1}{|c|}{$m_s$} &
\multicolumn{1}{|c|}{T} &
\multicolumn{1}{|c|}{$n_B$} &
\multicolumn{1}{|c|}{$\epsilon_d$} &
\multicolumn{1}{|c|}{$\epsilon_{dI}$} &
\multicolumn{1}{|c|}{$\epsilon_s$} &
\multicolumn{1}{|c|}{$\epsilon_{sI}$} \\
\multicolumn{1}{|c|}{} &
\multicolumn{1}{|c|}{(MeV)} &
\multicolumn{1}{|c|}{(MeV)} &
\multicolumn{1}{|c|}{$(fm^{-3})$} &
\multicolumn{1}{|c|}{$(erg/cm^{3}/s)$} &
\multicolumn{1}{|c|}{$(erg/cm^{3}/s)$} &
\multicolumn{1}{|c|}{$(erg/cm^{3}/s)$} &
\multicolumn{1}{|c|}{$(erg/cm^{3}/s)$} \\
\hline
   &     &   &0.6&4.34$\times 10^{+23}$&1.14$\times 10^{+25}$&1.69$\times
10^{+24}$&0.00$\times 10^{+00}$\\
   &     &0.1&1.0&2.84$\times 10^{+23}$&1.26$\times 10^{+25}$&2.18$\times
10^{+24}$&5.29$\times 10^{+24}$\\
   &     &   &1.4&2.09$\times 10^{+23}$&1.34$\times 10^{+25}$&2.40$\times
10^{+24}$&5.62$\times 10^{+24}$\\
\cline{3-8}
   &     &   &0.6&6.73$\times 10^{+25}$&4.68$\times 10^{+28}$&7.79$\times
10^{+27}$&0.00$\times 10^{+00}$\\
   &150.0&0.4&1.0&4.26$\times 10^{+25}$&5.16$\times 10^{+28}$&8.75$\times
10^{+27}$&2.17$\times 10^{+28}$\\
   &     &   &1.4&3.15$\times 10^{+25}$&5.48$\times 10^{+28}$&9.21$\times
10^{+27}$&2.30$\times 10^{+28}$\\
\cline{3-8}
    &     &   &0.6&3.01$\times 10^{+26}$&5.33$\times 10^{+29}$&8.79$\times
10^{+28}$&0.00$\times 10^{+00}$\\
    &     &0.6&1.0&2.01$\times 10^{+26}$&5.88$\times 10^{+29}$&9.46$\times
10^{+28}$&2.47$\times 10^{+29}$\\
0.05&     &   &1.4&1.56$\times 10^{+26}$&6.24$\times 10^{+29}$&9.64$\times
10^{+28}$&2.62$\times 10^{+29}$\\
\cline{2-8}
   &     &   &0.6&2.90$\times 10^{+24}$&2.27$\times 10^{+25}$&3.08$\times
10^{+23}$&0.00$\times 10^{+00}$\\
   &     &0.1&1.0&2.27$\times 10^{+24}$&2.53$\times 10^{+25}$&2.29$\times
10^{+24}$&0.00$\times 10^{+00}$\\
   &     &   &1.4&1.85$\times 10^{+24}$&2.70$\times 10^{+25}$&3.76$\times
10^{+24}$&0.00$\times 10^{+00}$\\
\cline{3-8}
   &     &   &0.6&6.70$\times 10^{+26}$&9.29$\times 10^{+28}$&1.09$\times
10^{+28}$&0.00$\times 10^{+00}$\\
   &200.0&0.4&1.0&4.37$\times 10^{+26}$&1.03$\times 10^{+29}$&1.58$\times
10^{+28}$&0.00$\times 10^{+00}$\\
   &     &   &1.4&3.24$\times 10^{+26}$&1.11$\times 10^{+29}$&1.81$\times
10^{+28}$&0.00$\times 10^{+00}$\\
\cline{3-8}
   &     &   &0.6&2.84$\times 10^{+27}$&1.06$\times 10^{+30}$&1.47$\times
10^{+29}$&0.00$\times 10^{+00}$\\
   &     &0.6&1.0&1.86$\times 10^{+27}$&1.18$\times 10^{+30}$&1.89$\times
10^{+29}$&0.00$\times 10^{+00}$\\
   &     &   &1.4&1.39$\times 10^{+27}$&1.26$\times 10^{+30}$&2.10$\times
10^{+29}$&0.00$\times 10^{+00}$\\
\hline
\end{tabular}
\end{center}
\newpage
\noindent Table 3.  Baryon number density $n_B$, Fermi momenta of u-quark
$p_{F}(u)$, d-quark $p_{F}(d)$ and electron $p_{F}(e)$ for
different $\alpha_c$, where $\Delta
p_d=p_{F}(u)+p_{F}(e)-p_{F}(d)$ .
\vspace {0.2in}
\begin{center}
\begin{tabular}{|c|c|c|c|c|c|}
\hline
\multicolumn{1}{|c|}{$\alpha_c$} &
\multicolumn{1}{|c|}{$n_B$} &
\multicolumn{1}{|c|}{$p_{F}(u)$} &
\multicolumn{1}{|c|}{$p_{F}(d)$} &
\multicolumn{1}{|c|}{$p_{F}(e)$} &
\multicolumn{1}{|c|}{$\Delta p_d$} \\
\multicolumn{1}{|c|}{} &
\multicolumn{1}{|c|}{($fm^{-3}$)} &
\multicolumn{1}{|c|}{(MeV)} &
\multicolumn{1}{|c|}{(MeV)} &
\multicolumn{1}{|c|}{(MeV)} &
\multicolumn{1}{|c|}{(MeV)} \\
\hline
   &0.60&357.80&449.20&99.15&7.75\\
0.1&1.00&424.22&532.58&117.56&9.20\\
   &1.40&474.57&595.79&131.51&10.29 \\
\hline
   &0.60&357.71&449.25&95.43&3.89\\
0.05&1.00&424.11&532.65&113.15&4.61\\
   &1.40&474.45&595.87&126.57&5.15\\
\hline
\end{tabular}
\end{center}
\newpage
\noindent Table 4. Baryon number density $n_B$, Fermi momenta of
u-quark $p_{F}(u)$, d-quark $p_{F}(d)$, s-quark $p_{F}(s)$ and
electron $p_{F}(e)$ for different $m_s$ and different $\alpha_c$
, where $\Delta p_d=p_{F}(u)+p_{F}(e)-p_{F}(d)$ and $\Delta
p_s=p_{F}(u)+p_{F}(e)-p_{F}(s)$
\vspace {0.2in}
\begin{center}
\begin{tabular}{|c|c|c|c|c|c|c|c|c|}
\hline
\multicolumn{1}{|c|}{$m_s$} &
\multicolumn{1}{|c|}{$\alpha_c$} &
\multicolumn{1}{|c|}{$n_B$} &
\multicolumn{1}{|c|}{$p_{F}(u)$} &
\multicolumn{1}{|c|}{$p_{F}(d)$} &
\multicolumn{1}{|c|}{$p_{F}(s)$} &
\multicolumn{1}{|c|}{$p_{F}(e)$}&
\multicolumn{1}{|c|}{$\Delta p_d$}&
\multicolumn{1}{|c|}{$\Delta p_s$} \\
\multicolumn{1}{|c|}{(MeV)} &
\multicolumn{1}{|c|}{} &
\multicolumn{1}{|c|}{($fm^{-3}$)} &
\multicolumn{1}{|c|}{(MeV)} &
\multicolumn{1}{|c|}{(MeV)} &
\multicolumn{1}{|c|}{(MeV)} &
\multicolumn{1}{|c|}{(MeV)}&
\multicolumn{1}{|c|}{(MeV)}&
\multicolumn{1}{|c|}{(MeV)} \\
\hline
     &   &0.60&356.99&360.06&353.86&3.33&0.26&6.46\\
     &0.1&1.00&423.26&424.81&421.69&1.69&0.14&3.26\\
150.0&   &1.40&473.49&474.27&472.72&0.84&0.06&1.61\\
\cline{2-9}
     &    &0.60&356.99&365.89&347.62&9.28&0.38&18.65\\
     &0.05&1.00&423.26&430.29&415.98&7.33&0.30&14.61\\
     &    &1.40&473.49&479.49&467.34&6.25&0.25&12.40\\
\hline
     &   &0.60&356.99&365.93&347.58&9.70&0.76&19.11\\
200.0&0.1&1.00&423.26&429.10&417.25&6.34&0.50&12.35\\
     &   &1.40&473.49&477.66&469.25&4.52&0.35&8.76\\
\cline{2-9}
     &    &0.60&356.99&374.26&337.85&18.00&0.73&37.14\\
     &0.05&1.00&423.26&437.14&408.39&14.47&0.59&29.34\\
     &   &1.40&473.49&485.45&460.90&12.46&0.50&25.05\\
\hline
\end{tabular}
\end{center}
\end{document}